\documentclass{article}

\usepackage{arxiv}

\usepackage[utf8]{inputenc} 
\usepackage[T1]{fontenc}    
\usepackage{hyperref}       
\usepackage{url}            
\usepackage{booktabs}       
\usepackage{amsfonts}       
\usepackage{nicefrac}       
\usepackage{microtype}      
\usepackage{lipsum}
\usepackage{graphicx}

\usepackage[numbers]{natbib}
\usepackage{amsmath, amssymb}

\usepackage[ruled,vlined]{algorithm2e}
\usepackage{multicol, multirow, booktabs}
\usepackage{array,ragged2e}
\usepackage{tikz}
\usepackage{pgfplots}
\pgfplotsset{width=7.5cm,compat=1.12}
\usepgfplotslibrary{fillbetween}

\SetKwRepeat{Do}{do}{while}

\newcommand{\axisStyle}[2]{title style={at={(0.5,1.1)},anchor=north, yshift=-0.1},
    title=#1,
  x tick label style={/pgf/number format/fixed},
  ymajorgrids,
  ymin=0.0,
  ymax=1.0,
  ytick={0.0,0.1,0.2,0.3,0.4,0.5,0.6,0.7,0.8,0.9,1.0},  
    width=8.5cm,
    height=7cm,
  ylabel=#2 robustness,
  xlabel=Size,
    enlarge y limits={0.1,upper},
    enlarge x limits=0.15,    
    legend pos=south east,
  ybar=1.4pt,
    bar width=10pt,}

\newcommand{\bools}{\ensuremath{\mathbb{B}}}

\newcommand{\bwd}{\ensuremath{\textsc{Bwd}}}
\newcommand{\fwdClosed}{\ensuremath{\textsc{Trap}}}
\newcommand{\var}[1]{\texttt{#1}}
\newcommand{\pre}{\textsc{Pre}}
\newcommand{\post}{\textsc{Post}}

\newcommand{\async}{\ensuremath{\textsc{STG}}}

\newcolumntype{L}[1]{>{\raggedright\let\newline\\\arraybackslash\hspace{0pt}}m{#1}}
\newcolumntype{C}[1]{>{\centering\let\newline\\\arraybackslash\hspace{0pt}}m{#1}}
\newcolumntype{R}[1]{>{\raggedleft\let\newline\\\arraybackslash\hspace{0pt}}m{#1}}

\title{Robust Control of Partially Specified Boolean Networks}

\author{
 Luboš Brim \\
  Faculty of Informatics\\
  Masaryk University\\
  \texttt{brim@fi.muni.cz} \\
   \And
 Samuel Pastva \\
  Faculty of Informatics\\
  Masaryk University\\
  \texttt{xpastva@fi.muni.cz} \\
  \And
 David Šafránek \\
  Faculty of Informatics\\
  Masaryk University\\
  \texttt{xsafran1@fi.muni.cz} \\
  \And
 Eva Šmijáková \\
  Faculty of Informatics\\
  Masaryk University\\
  \texttt{xsmijak1@fi.muni.cz} \\
}

\begin{document}
\maketitle

\begin{abstract}
Regulatory networks (RNs) are a well-accepted modelling formalism
in computational systems biology. The control of RNs is currently receiving a lot of attention because it provides a theoretical and computational basis for cell reprogramming -- an attractive technology developed in regenerative medicine. By solving the control problem, we learn which parts of a biological system should be perturbed in order for it to stabilise the system in the desired phenotype. 

In this paper, we use \emph{Boolean networks (BNs) with asynchronous update} to represent the dynamics of an RN as a discrete finite-state system. Furthermore, we allow the specification of the Boolean model representing a given RN to be incomplete (partial). This is crucial in cases where the exact Boolean update functions of the BN are not fully known (e.g. due to a lack of measured data). To that end, we utilise the formalism of \emph{partially specified Boolean networks} which allows us to cover every possible behaviour reflecting the unspecified parts of the system. 
Such an approach inevitably causes a significant state explosion. This problem is efficiently addressed by using \emph{symbolic methods} to represent both the unspecified model behaviour as well as all possible perturbations of the system. Our framework supports several ways to control the system by employing one-step, temporary, and permanent perturbations.

Additionally, to make the control design efficient and practically applicable, the optimal control should be minimal in terms of the size -- the number of perturbed system components. Moreover, in a partially specified model, a control may achieve the desired stabilisation only for a subset of the possible fully specified instantiations of the model. To address these aspects in the control design, we utilise several quantitative measures. In particular, apart from the \emph{size} of perturbation necessary to achieve control, we also examine its \emph{robustness} -- a portion of fully specified models for which the control is applicable.

We show that the proposed symbolic methods solving the control problem for partially specified BNs are efficient and scale well with the number of unspecified model elements. The provided experiments demonstrate that our algorithms can handle relatively large BNs. We also evaluate the robustness metrics in cases of all three studied control types. The robustness metric tells us how big a proportion of fully defined systems the given perturbation works. Our experiments support the hypothesis that one-step perturbations may be less robust than temporary or permanent perturbations.

This is a full version of a paper that is submitted to a journal.



\end{abstract}




\section{Introduction}
\label{s:introduction}

The control theory is on the rise in computational systems biology thanks to its practical potential~\cite{Kitano2002,richardson2015,borriello2021}. In particular, the solution to the control problem for an \emph{in silico} model of a cell can provide the fundamental basis for experimental design for reprogramming of the cell \emph{in vitro}~\cite{Cherry2012}.
The common applications of cell reprogramming are differentiation of T-cells~\cite{Miskov-Zivanovra97,Saadatpour2011,Zanudo2015}, reprogramming of embryonic stem cells~\cite{Herberts2011,Singh2009,Takahashi2006,Young2011,Wernig2007}, curing obesity by converting fat cells to the brown type~\cite{zhu2015}, and recovering malfunctioning cells~\cite{motter2008} or even organs~\cite{goligorsky2019}. 


\emph{Regulatory network} (RN) is a widely used model for \emph{in silico} experiments because it allows capturing numerous relationships in complex biochemical systems~\cite{davidson2005, Hecker2009}. Nonetheless, RNs represent a highly abstract level of view on a system, and more specific (mechanistic) model adaptations are usually employed for realising computational tasks such as a simulation of the system dynamics. Boolean models of RNs~\cite{Barbuti2020,albert2004}, or Boolean networks (BNs) for short, are often employed to study the RN dynamics. In BNs, the dynamical behaviour is specified by a set of Boolean update functions defining the evolution of variables over time. In this paper, we focus on BNs with the asynchronous update that reflects well the concurrent dynamics of individual components (genes) in the RN~\cite{thomas1995dynamical,Schwab2020}. This updating scheme results in a non-deterministic state-transition graph (STG), as opposed to a simpler, deterministic graph which arises under the less realistic synchronous scheme~\cite{Aracena2009}.



A critical problem of BN modelling is to fully determine the update functions from data or literature~\cite{berestovsky2013evaluation}. In many cases, the knowledge of update functions is incomplete, and the modellers must address this problem~\cite{Grieco2013}. Other possible situations when multiple model adaptations should be considered at the same time are differences between intrinsic and extrinsic observations~\cite{Geris2016}, or structural changes in RNs occurring over time~\cite{Martin2016}. We address these issues by targeting analysis of \emph{partially specified} Boolean models. We represent the partial specification in BN by allowing the use of \emph{uninterpreted functions} as part of BN's update functions. Intuitively, such an uninterpreted function stands for some unknown but fixed part of the network's dynamics. By fixing the behaviour of these uninterpreted functions, we can then obtain a wide range of standard (i.e. fully specified) BNs, which we refer to as \emph{instantiations}. Technically, in our framework a partially specified BN in represented as a BN with inputs.

The goal of \emph{Boolean network control} is to influence the behaviour of the network so that it stabilises in a given \emph{attractor} -- a minimal component of the system, which once reached, cannot be escaped. A typical way to influence (change) the BN's behaviour is to perturb the values of some of its variables. Since the practical realisation of perturbations requires non-trivial effort, we typically require the number of perturbations to be as small as possible. To that end, the control is usually optimised by minimising the number of controlled variables~\cite{Baudin2019}. 

However, the second factor that needs consideration when the network is not fully known is that a perturbation may succeed only for some fraction of instantiations of the partially specified network. Meanwhile, for other instances, the target attractor may not be reached. We quantify this aspect using a \emph{robustness} metric -- a portion of fully specified model instantiations for which the resulting perturbation can control the respective network to achieve the desired attractor.

There are multiple possible ways a perturbation can be applied to a biological system. In this paper, we focus on \emph{temporary} and \emph{permanent} perturbations. As such, we consider that the value of perturbed variables changes simultaneously and is held fixed for multiple time-steps. In the temporary case, the system is eventually left to behave as unperturbed. In contrast, the variables are fixed ad infinitum during permanent perturbations.

There is a large body of existing work on the BN control problem. 
Most commonly, the \emph{source-target} variant of the BN control problem is considered, where both the source and the target state (or attractor) are specified. Other types of control include target control~\cite{Kim2013} (the controlled system must reach the target attractor regardless of the initial state) and full-control~\cite{Fiedler2013} (a control strategy between all attractor pairs is desired).

In the case of BNs utilising the synchronous update, the problem is far less complicated, with efficient approaches already shown in~\cite{Kim2013,Zhao2015,Moradi2019}. The main reason is that in a synchronous BN, due to the determinism of the network, its attractors are simpler (only single-state and cyclic attractors appear), and from each initial state, only one attractor is reachable. Meanwhile, an asynchronous network can contain larger, more complex attractors, and multiple attractors can be reachable from any network state. Owing to the more detailed mechanistic view of the concurrent regulatory events~\cite{Schwab2020}, control of asynchronous BNs is currently receiving a lot of attention. For example, the approach developed in~\cite{Zanudo2015,Rozum2021a,Rozum2021b} computes a set of relevant BN variables based on the identification of
particular motifs in the RN. However, perturbations based on such stable motifs control the BNs only when applied permanently. 

In some cases, such permanent perturbations may not be feasible~\cite{Cornelius2013}. To that end, the concept of a \emph{one-step} perturbation has been developed in~\cite{Baudin2019}. This type of perturbation is applied in the initial state, and after that,
the system evolves according to its original dynamics. This approach has motivated a branch of methods based on the identification of the \emph{strong basin} of the target attractor (i.e. the states which \emph{must} eventually reach said attractor). Subsequently, the concept of temporary perturbations was introduced and studied in~\cite{MandonHP17,Su2019FM,JunPang2020a}, filling the gap between one-step and permanent perturbations. Temporary perturbations were shown to be very efficient: While the perturbation does not need to be enforced indefinitely, the number of perturbed variables necessary to control the network is usually minimal compared to a one-step perturbation. Even more complex control strategies, considering temporal perturbation sequences are studied in~\cite{Jaoude16,Mandon2019CMSB,Pardo19,Su2020}. The tool CABEAN comprises these methods~\cite{Su2020Bioinformatics}.

Another line of research addressing temporary perturbations is based
on the so-called trap-spaces~\cite{Fontanals2020}. In this approach, the target attractor is also only partially known. The developed
method allows the specification of the target via a phenotype -- a subset of desired variable values, instead of an attractor. Such phenotype control is explored in more detail in~\cite{Fontanals2021}, where a method based on model checking is developed. However, the scalability of this method may be limited by the necessity to scan through the available perturbations.

All results mentioned so far have been developed only for fully specified BNs. It is worth noting that the incomplete specification is far more computationally demanding. Each possible full model specification generates a new BN with a unique STG. Therefore, the \emph{state-space explosion} (state-space size is exponential in the number of BN variables) is multiplied by the number of possible BN instances. In the worst case, this number is doubly-exponential with respect to the number of inputs of the network update functions~\cite{Wang2012}. The size of the resulting multi-graph makes control algorithms developed for fully specified asynchronous BNs (by computing the control for each BN instance distinctly) intractable. Consequently, addressing the control problem in partially specified BNs requires substantially novel algorithms.



In~\cite{Brim2021}, the control of partially specified BNs have been
addressed for the first time. In particular, the paper introduced a method for one-step
source-target control of BNs with logical parameters (encoding the partial specification of BNs). On the technical side, the method has been supported by a semi-symbolic algorithm. In particular, the unspecified parts are
represented symbolically using binary decision diagrams (BDDs)~\cite{Bryant86} while the STG is explicit~\cite{Benes2019}. The algorithm is based on the identification of the attractor's strong basin for all parametrisations simultaneously. 

In this paper, the methodology and the algorithmics for the control of partially specified BNs are significantly advanced. First, we extend the semi-symbolic approach to an entirely symbolic graph representation. This allows processing models with a higher number of variables. Second, we develop new algorithms for more permissive types of perturbations (in terms of minimal size and robustness) -- temporary and permanent. Last but not least, the novelty of our approach lies in a new method we use to simulate the perturbed BN. Instead of constructing a new graph for each perturbation, we model all perturbations in a single transition-labelled graph. We show that our approach scales well for highly unspecific models and outperforms the semi-symbolic approach from~\cite{Brim2021} primarily in terms of processable model size. On biologically relevant models, we also demonstrate that temporary and permanent perturbations are more robust compared to available one-step perturbations.

\section{Theory}

This section presents a formal introduction to the topics discussed in this paper. Mainly, it introduces the modelling framework of partially specified Boolean networks and the control problems for this type of model.

At first, let us define $\bools = \{ 0, 1\}$ to be the set of Boolean values, and $\bools_* = \{ 0, 1, * \}$ to be an extension of $\bools$ which also admits a \emph{free} value $*$ (i.e. neither $\mathit{true}$ nor $\mathit{false}$). Additionally, we write $\bools^n$ to denote the set of all $n$-element vectors over $\bools$. For each vector $x \in \bools^n$, by convention, $x_i$ is the $i$-th element and $x[i \mapsto b]$ is the copy of $x$ with the $i$-th element set to $b \in \bools$.

\textbf{Asynchronous Boolean networks:} As the name suggests, a Boolean network (BN) assumes $n$ Boolean \emph{variables}, whose state evolves over time. The state of such a network is therefore an $n$-element vector $s \in \bools^n$ (we say that $\bools^n$ is the \emph{state space} of the BN). Additionally, we assume that the network can have $m$ Boolean \emph{inputs} (sometimes also referred to as \emph{constants}), which have arbitrary but fixed values. Consequently, we say that $\bools^m$ is the input space of the BN and $c \in \bools^m$ is an input valuation. Finally, the Boolean network $B$ itself is a collection of update functions $B = (b_1, \ldots, b_n)$, such that each $b_i: \bools^n \times \bools^m \to \bools$ takes the current network state and input valuation, and outputs a new value for the $i$-th variable. Typically, if function $b_i$ depends on the value of variable $j$, we say that $j$ \emph{regulates} $i$. These relationships are prescribed by a directed \emph{influence graph} (also called a \emph{regulatory network}) which can serve as a visual representation of the BN structure. Note that the inputs and variables are not explicitly distinguished in some literature since an input can be simply seen as a variable with identity as its update function.

We assume that the variables of the network are updated \emph{asynchronously}, i.e. each state-transition updates exactly one variable based on the output of $b_i$. Consequently, we can describe the dynamics of a BN with inputs using a \emph{coloured state-transition graph}. A coloured graph $\async(B) = (C, V, E)$ consists of three elements. A finite set of \emph{colours} $C = \bools^m$ corresponding to the input valuations of the network. A finite set of \emph{vertices} $V = \bools^n$ corresponding to the state space of the network. And finally, the edges $E \subseteq V \times C \times V$ are constructed based on the update functions using the asynchronicity assumption explained above:
\begin{align*}
      (s, c, t) \in E \Leftrightarrow \exists i \in [1,n]. (t = s[i \mapsto b_i(s, c)] \land s \not= t)
\end{align*}

Additionally, to guarantee that $E$ is total for every $c$, we assume that a self-loop $(s, c, s) \in E$ is present for a state $s$ that has no other outgoing edges. We write $s \xrightarrow{c} t$ when $(s,c,t) \in E$, and $s \xrightarrow{c}^* t$ when $(s, c, t) \in E^*$ (i.e. the reflexive and transitive closure of $E$). Intuitively, the coloured graph $\async(B)$ can be seen as a family of standard directed graphs over a common set of vertices, where the choice of input valuation determines the edges of the graph. Because all variables and inputs are Boolean, this structure has a fairly straightforward symbolic representation in terms of binary decision diagrams, as we later demonstrate.

Additionally, for any coloured graph we can define a \emph{run} $\pi$ under colour $c$ as a maximal sequence of states $s, s_1, \ldots$ such that $(s_i, c, s_{i+1}) \in E$ for every $i$. Note that such $\pi$ is necessarily infinite due to $E$ being total. State $s_0$ is the \emph{initial state} of such a run. State $s_0$ is the \emph{initial state} of such a run. We write $\pi(i)$ to denote the $i$-th state $s_i$ of $\pi$, and $inf(\pi)$ to denote all states which appear in $\pi$ infinitely often.

Since Boolean networks primarily function as models of biological systems which have some inherent (but in our case unquantified) stochasticity, we typically assume that the system can only exhibit \emph{fair} runs. That is, available transitions cannot be delayed indefinitely~\cite{abadi1995}. To capture this property, we write $\Pi_c(s_0)$ to denote the set of all fair runs under colour $c$ which originate in the state $s_0$.

\textbf{Partially specified Boolean networks:} A Boolean network with inputs allows us to easily encode a wide range of biochemical systems in a machine-friendly format. However, for systems with a high degree of uncertainty, it often fails to capture this uncertainty in a way that is understandable to a human reader.

To address this issue, the actual input of our method are \emph{partially specified} Boolean networks, which explicitly allow parts of the update functions to be designated as unknown. Assume that $f_1^{(a_1)}$, $f_2^{(a_2)}$, $\ldots$ are symbols standing in for some uninterpreted (fixed but arbitrary) Boolean functions ($a_i$ being the function arity). A partially specified BN then assumes $n$ variables and $p$ uninterpreted functions. In this kind of network, every update function $b'_i$ is specified as a Boolean expression that can contain the function symbols $f_1, \ldots, f_p$.

This formalism is often easier to comprehend, as the uncertainty in dynamics is tied to the update functions instead of inputs. However, it is also not immediately clear how such a network should be represented symbolically. It is relatively easy to translate a partially specified network into a~BN with inputs, though. Any uninterpreted function $f_i^{(a)}$ can be encoded in terms of $2^a$ fresh Boolean inputs if we consider that $j$-th input denotes the output of $f_i^{(a)}$ in the $j$-th row of its function table. Formally, this translation can be achieved using a~repeated application of the following expansion rule:
\begin{align*}
      f(\alpha_1, \ldots, \alpha_a) \equiv (\alpha_1 \Rightarrow f'_1(\alpha_2, \ldots, \alpha_a)) \land (\neg\alpha_1 \Rightarrow f'_2(\alpha_2, \ldots, \alpha_a))
\end{align*}
Here, $f'_1$ and $f'_2$ are fresh uninterpreted functions of arity $a-1$, and $\alpha_i$ are arbitrary Boolean expressions. Using this rule, we can always convert a partially specified network to a Boolean network with inputs. Nonetheless, the number of inputs will be exponential with respect to the arity of the employed uninterpreted functions (since each application of the rule replaces one uninterpreted function with two fresh ones).

\textbf{Network attractors and source-target control:} To understand the dynamics of a BN (or a partially specified BN translated to a BN with inputs), we have to be familiar with the notion of \emph{attractor}. Informally, an attractor is the smallest set of states that cannot be escaped and hence the system eventually converges into it. Formally, an attractor $A \subseteq V$ for colour $c \in C$ is the smallest set such that for every $s \in A$, if $s \xrightarrow{c} t$ for some $t \in V$, then $t \in A$. Note that for every fair run $\pi$, $inf(\pi)$ is always some attractor of the network. In Boolean networks, different attractors typically correspond to different real-world phenotypes which the system can exhibit.

Therefore, for a real-world Boolean network, one of the fundamental questions is whether (and if so, how, or even how efficiently) the network can be controlled into a specific attractor. We assume that the network is controlled by a \emph{perturbation} which fixes the values of certain variables to constants. Formally, a perturbation is a vector $q \in \bools_*^n$, where for every $q_i = *$, we say that the $i$-th variable is unperturbed, whereas for $q_i = 0$ or $q_i = 1$, we say that the variable is perturbed to either $0$ or $1$. For a state $s \in \bools^n$, we then write $q(s)$ to denote a copy of $s$ where the perturbation $q$ has been applied. Formally, $q(s)_i = s_i$ when $q_i = *$, and $q(s)_i = q_i$ otherwise. Finally, we overload the notation $\Pi_c^q(s)$ to denote the set of fair runs under colour $c$ starting in state $s$, where the runs are restricted to the edges of $\async(B)$ that only update the \emph{unperturbed} variables in $q$. For the sake of maintaining the property that every fair run is infinite, we assume that this restriction admits a self-loop on every state that cannot update any unperturbed variable (i.e. any state that would otherwise have no outgoing transitions). That is, intuitively, along the run in $\Pi_c^q(s)$, the perturbed variables are never updated, and if a state from which no outgoing edges using unperturbed variable exist, the state is repeated forever.

While the notion of a perturbation should now be clear, the exact temporal nature of the perturbation can be still different based on the biological assumptions of the perturbed system and the real-world perturbation method. In every instance, a system in state $s$ is perturbed into the state $q(s)$ by applying the perturbation $q$. However, we recognise three types of state perturbations based on the subsequent evolution of the system: (a) \emph{one-step} perturbation: after perturbation, all system variables again evolve freely; (b) \emph{permanent} perturbation: once perturbed, only unperturbed variables can evolve. Perturbed variables are fixed to their respective constant values; (c) \emph{temporary} perturbation: after perturbation, the perturbed variables remain constant for some arbitrary but finite number of time-steps but are eventually allowed to evolve freely again.

Based on these three types of perturbations, we can then formalise three control problems. In every case, we assume a source state $source \in \bools^n$ together with a target state $target \in \bools^n$, and a fixed input valuation $c \in \bools^m$:
\begin{itemize}
      \item One-step control: Identify every perturbation $q \in \bools_*^n$ s.t. for every $\pi \in \Pi_c(q(source))$, $target \in inf(\pi)$. That is, after performing the perturbation, $target$ is always visited infinitely often without further restricting the network in any way.
      \item Permanent control: Identify every perturbation $q \in \bools_*^n$ s.t. for every $\pi \in \Pi_c^q(q(source))$, $target \in inf(\pi)$. That is, $target$ is always visited infinitely often assuming the perturbed variables remain constant.
      \item Temporary control: Identify every perturbation $q \in \bools_*^n$ s.t. for every $\pi \in \Pi_c^q(q(source))$, there is some $j \in \mathbb{N}$ where for every $\pi' \in \Pi_c(\pi(j))$, $target \in inf(\pi')$. Intuitively, the system can evolve for an arbitrary number of time steps while assuming the perturbed variables are constant, but eventually, every such run reaches a state from which the system can evolve into $target$ as if unperturbed.
\end{itemize}

In the case of one-step and temporary control, such a perturbation only exists when $target$ is a member of some attractor of $B$ (or perturbed $B$ in case of permanent control). Otherwise, no fair run will visit it infinitely often, regardless of applied perturbations. It is therefore typically reasonable to only consider input valuations for which $target$ is indeed an attractor state~\cite{Brim2021}. On the contrary, the permanent perturbation may induce the appearance of new attractors, including attractors containing the $target$ state. Nonetheless, this scenario is solving a much harder biological task - introduction of a new phenotype (phenotypic variation)~\cite{Forsman2014}. We leave this line of research for future work and consider only colours containing an attractor with the $target$ state.

Ideally, we would also like to select for perturbations that are biologically feasible (typically, due to physical limitations, not every variable can be perturbed) and that perturb the smallest number of variables (to minimise the cost and complexity of a perturbation). However, the validity of a perturbation is also tied to the associated input valuation $c$, which may describe some unknown parts of the system that we cannot influence. As such, most perturbations will correctly control the network only for some subset of input valuations. Depending on how large this subset is, we can then talk about \emph{robustness} $\rho$ of a one-step, permanent, or temporary perturbation $q$ with respect to the network inputs~\cite{Brim2021}:
\begin{align*}
\centering
      \rho(q) = \frac{|\{ c \in \bools^m \mid q \text{ controls } B \text{ under } c \}|}{|\{ c \in \bools^m \mid target \text{ is an attractor state} \}|}
 \end{align*}

Robustness is a critical factor when selecting viable perturbations in BNs with inputs. To compute this property, our goal is to compute a \emph{control relation} $\mathcal{C}_? \subseteq \bools^n_* \times \bools^m$ (here, $?$ is either $O$ for one-step, $P$ for permanent, or $T$ for temporary). This relation includes all pairs $(q, c) \in \bools^n_* \times \bools^m$ such that $q$ controls $B$ under input valuation $c$ from $source$ to $target$ assuming one-step ($\mathcal{C}_O$), permanent ($\mathcal{C}_P$), or temporary ($\mathcal{C}_T$) control.

Our robustness measure gives us the ability to compare the ``quality'' of particular perturbations. In order to also evaluate the ``quality'' of different control relations (or control approaches), we extend the notion of robustness to be measurable on a set of perturbations. Specifically, given a set of perturbations $Q$, we define \emph{maximal robustness} $\rho_{max}(Q)$ to be the value of maximal robustness within the given set, and \emph{union robustness} $\rho_{union}(Q)$ to be the union of all input valuations controlled by the given set with respect to the relevant colours: 

$$ \rho_{max}(Q) = max\{\rho(q) \mid q \in Q\} $$ 
$$    
      \rho_{union}(Q) = \frac{|\bigcup_{q\in Q}\{ c \in \bools^m \mid q \text{ controls } B \text{ under } c \}|}{|\{ c \in \bools^m \mid target \text{ is an attractor state} \}|}
$$

Intuitively, the maximal robustness $\rho_{max}$ gives us the best robustness that can be achieved within a particular set using a single perturbation. Meanwhile, union robustness $\rho_{union}$ is the measure of how well a system can be controlled \emph{overall} using perturbations from $Q$. Both measures are worth exploring, as each represents a slightly different optimisation goal: While a set $Q_1$ with a high $\rho_{max}$ is generally more likely to provide a practically viable control strategy, a set $Q_2$ with a high $\rho_{union}$ may achieve control even for cases that are not feasible using $Q_1$.

\section{Methods}

Real-world Boolean networks can have a large number of inputs and variables (and consequently, many admissible perturbations). It is not possible to explore the STGs of such networks explicitly. We propose several symbolic algorithms to address this issue. These are based on the existing symbolic approaches to control of BNs \emph{without} inputs (i.e. networks with complete knowledge). However, the addition of inputs introduces extra complexity, which our methods need to address. In particular, identifying a single minimal perturbation is not sufficient, as the robustness of such a perturbation can be very low, making it unlikely to work in practice. To provide a reasonably reliable method, one needs to consider a large set of perturbations. Consequently, for performance reasons, our algorithms cannot test only a single perturbation at a time but need to explore the whole perturbation space symbolically.

\textbf{Symbolic computation model:} We employ symbolic representation using binary decision diagrams (BDDs)~\cite{Bryant86}. In this representation, a BDD corresponding to a Boolean formula $\varphi$ is used to represent a set (or a relation) $X$ of Boolean vectors where the elements of $X$ are exactly the satisfying valuations of $\varphi$. Consequently, any subset of vertices or colours of $\async(B)$, i.e. $V = \bools^n$ and $C = \bools^m$, or any relation $R \subseteq V \times C$ can be represented in this way. To represent sets of perturbed variables, we will also refer to a set $P = \bools^n$. But note that the members of this set are not perturbations per se (i.e. members of $\bools_*^n$). For each $p \in P$, value $p_i = 1$ indicates only that the variable $i$ is perturbed, not the perturbation value itself.

The reasoning behind this choice of encoding is that after the network is perturbed, the perturbation values are contained within the state of the network variables. To get the full picture of the network's state under multiple different perturbations and input valuations, we can thus use a relation $X \subseteq V \times C \times P$. For $(s, c, p) \in X$, $p_i = 1$ indicates that, assuming input valuation $c$, $i$-th variable is perturbed, and $s_i$ gives the actual value of the perturbation (for $p_i = 0$, $s_i$ is the normal mutable state of the network). Finally, we use $\mathcal{S} = V \times C \times P$ as a shorthand for this extended set of perturbable states and input valuations. We also write that a pair $(s, p) \in V \times P$ is \emph{equivalent} to a perturbation $q \in \bools_*^n$, $(s, p) \equiv q$, if $q_i = *$ when $p_i = 0$, and $q_i = s_i$ when $p_i = 1$. Observe that this approach is more efficient than simply incorporating the set of all possible perturbations $\bools^n_*$ directly into the encoding. Each $\bools_*$ would require more than one bit per network variable to encode, wherease our approach only adds one bit on top of the existing coloured state space.

On these symbolic sets and relations, we can perform standard set operations ($\cap, \cup, \setminus, \times$, etc.) as these correspond to logical operations on BDDs. Additionally, we use $\exists_{P}(X \subseteq \mathcal{S})$ to denote existential projection over the perturbed variables, i.e. $\exists_{P}(X) = \{ (s, c) \mid \exists p \in P. (s, c, p) \in X \}$. This can also be implemented using a single BDD operation. Finally, we need a mechanism to symbolically explore the edges of the STG $\async(B)$, but we also need to consider variable perturbations in this process. To do so, we define the two following symbolic operations:
\begin{align*}
      \pre(X \subseteq \mathcal{S}) &= \{ (s, c, p) \in \mathcal{S} \mid \exists t \in V.\ \exists i \in [1,n].\ t = s[i \mapsto b_i(s, c)] \land s \not= t \land p_i = 0 \}\\
      \post(X \subseteq \mathcal{S}) &= \{ (t, c, p) \in \mathcal{S} \mid \exists s \in V.\ \exists i \in [1,n].\ t = s[i \mapsto b_i(s, c)] \land s \not= t \land p_i = 0 \}\\     
\end{align*}
Recall that this is almost exactly the condition under which $(s, c, t) \in E$ of $\async(B)$. As such, $\pre$ and $\post$ compute the sets of predecessors and successors of the states in $X$ under their respective input valuations. However, for each computed transition, we also require that $p_i = 0$, i.e. the $i$-th variable is \emph{not} perturbed. Hence values of the perturbed variables remain constant, and in case $p_i = 0$ for all $i$, we get exactly the behaviour of $\async(B)$ without any perturbations. These two operations can also be implemented using BDDs, since $b_i$ is a Boolean function (which itself trivially corresponds to a BDD), and the rest are either logical operations or quantifications that can be realised within the domain of our BDD representation.

\textbf{Utility algorithms:} Our algorithms rely on the utility procedures defined in Algorithm~\ref{algo:sb}. Here, $\bwd$ is a standard backward reachability procedure, albeit over members of our extended state space $\mathcal{S}$. Meanwhile, the result of $\fwdClosed$ is a so-called \emph{trap set}~\cite{Fontanals2020}: a subset of $X$ from which one cannot escape $X$. Note that if $X$ contains some attractors of $B$, $\fwdClosed(X)$ will contain exactly those attractors and states from $X$ that can only reach these attractors. Meanwhile, if $X$ does not contain any attractor, $\fwdClosed(X)$ will be empty. 

Finally, note that while our $\post$ and $\pre$ procedures \emph{respect} the setting of perturbed variables within subsets of $\mathcal{S}$, they do not actually perform any perturbations (i.e. they do not \emph{force} perturbed variables to their perturbed values). To work around this issue, Algorithm~\ref{algo:sb} also describes a $\textsc{CanPerturb}$ function. This function takes a \emph{source} state, and a set of \emph{target} states, and computes the subset of \emph{target} states that can be reached from the \emph{source} via a perturbation:

To compute this information, we notice that the elements of \var{targets} not belonging to $\textsc{CanPerturb}(\var{source}, \var{target})$ are based on exactly those combinations of $(s, p) \in V \times P$ for which there is at least one variable $i$ where $s_i$ does not match $\var{source}_i$, but $p_i = 0$ (variable is not perturbed). That is, a ``perturbation step'' cannot change a value of an unperturbed variable: all values in $\var{source}$ are either perturbed, in which case a perturbation will update their value to match $s$, or unperturbed, in which case they must already agree with $s$. We thus go through all variables and remove all cases where this situation occurs using symbolic operations.

    \begin{algorithm}
      \SetKwProg{Fn}{Fn}{}{}
      \Fn{\upshape \bwd$(X \subseteq \mathcal{S})$}{
        \lRepeat{\upshape $X \textbf{ reaches fixpoint}$}{
          $X \gets X \cup \pre(X)$
        }
        \Return{$X$}\;
      } 
      \Fn{\upshape \fwdClosed$(X \subseteq \mathcal{S})$}{
        \tcc{Eliminate elements that escape from $X$ in one step.}
        \lRepeat{\upshape $X \textbf{ reaches fixpoint}$}{
          $X \gets X \setminus \pre(\post(X) \setminus X)$
        }
        \Return{$X$}\;
      }
      \Fn{\upshape \textsc{CanPerturb}$(\var{source} \in V, \var{targets} \subseteq \mathcal{S})$}{
        \For{$i \in \{ 1 \ldots n \}$}{
          $\var{diff}_i \gets \{ s \in V \mid s_i \not= \var{source}_i \}$\; 
          $\var{zero}_i \gets \{ p \in P \mid p_i = 0 \}$\;
          $\var{targets} \gets \var{targets} \setminus (\var{diff}_i \times C \times \var{zero}_i))$\;
        }
        \Return{\upshape \var{targets}}\;
      }
    \caption{Basic graph algorithms.}
    \label{algo:sb}
    \end{algorithm}    

    \begin{algorithm}
      \SetKwProg{Fn}{Fn}{}{}
      \Fn{\upshape \textsc{Permanent}$(\var{source} \in V, \var{target} \in V)$}{
        $\var{basin} \gets \fwdClosed(\bwd(\{ \var{target} \} \times C \times P))$\;
        $\var{map} \gets \textsc{CanPerturb}(\var{source}, \var{basin})$\;
        \Return{\upshape $(q, c) \in \mathcal{C}_{P} \Leftrightarrow (s,c,p) \in \var{map} \land (s,p) \equiv q$}\;
      }
      \Fn{\upshape \textsc{OneStep}$(\var{source} \in V, \var{target} \in V)$}{
        $\var{basin} \gets \fwdClosed(\bwd(\{ \var{target} \} \times C \times \{ 0 \}^n))$\;
        $\var{basin} \gets \exists_P(\var{basin}) \times P$\;
        $\var{map} \gets \textsc{CanPerturb}(\var{source}, \var{basin})$\;
        \Return{\upshape $(q, c) \in \mathcal{C}_{O} \Leftrightarrow (s,c,p) \in \var{map} \land (s,p) \equiv q$}\;
      }
      \Fn{\upshape \textsc{Temporary}$(\var{source} \in V, \var{target} \in V)$}{
        $\var{basin} \gets \fwdClosed(\bwd(\{ \var{target} \} \times C \times \{ 0 \}^n))$\;
        $\var{basin} \gets \exists_P(\var{basin}) \times P$\;
        $\var{basin} \gets \fwdClosed(\bwd(\var{basin}))$\;
        $\var{map} \gets \textsc{CanPerturb}(\var{source}, \var{basin})$\;
        \Return{\upshape $(q, c) \in \mathcal{C}_{T} \Leftrightarrow (s,c,p) \in \var{map} \land (s,p) \equiv q$}\;
      }     
      \caption{Control procedures.}
      \label{algo:permanent}
    \end{algorithm}  

\textbf{Control procedures:} Algorithm~\ref{algo:permanent} then describes bottom-up procedures which compute the relations $\mathcal{C}_O$, $\mathcal{C}_P$, and $\mathcal{C}_T$. In the case of \textsc{Permanent} control, the variable \var{basin} contains all combinations of states, input valuations and perturbed variables from which \var{target} is visited infinitely often on fair paths. Specifically, \textsc{Bwd} finds all states that can reach \var{target} using some fair run. Out of these, \textsc{Trap} eliminates the ones that may not always reach \var{target} but have the opportunity to also escape into other attractors. In particular, this also completely eliminates input valuations where \var{target} is not a part of any attractor. Finally, \textsc{CanPerturb} is used to only retain combinations that can be reached from \var{source} via a perturbation.

\textsc{OneStep} relies on a very similar principle, however, the initial computation of \var{basin} is happening without any perturbations enabled: $P$ is exchanged for a singleton set $\{ 0 \}^n$. This adheres to the intuition that in the one-step control problem, the evolution after the perturbation is not constrained. However, once \var{basin} is computed, we use $\exists_{P}$ to substitute $\{ 0 \}^n$ for $P$. This effectively ``enables'' all possible perturbations and procedure \textsc{CanPerturb} can then assesses which of such perturbations are feasible from the \var{source} state.

Finally, \textsc{Temporary} is conceptually a combination of both approaches: first, a \var{basin} is computed with all perturbations disabled. Then, $\{ 0 \}^n$ is exchanged for $P$, which enables perturbations. Afterwards, the combination of \textsc{Bwd} and \textsc{Trap} procedures is repeated, yielding a larger set of states that eventually always reach \var{target} only if the corresponding perturbation is applied temporarily. Finally, \textsc{CanPerturb} again limits this set to perturbations that are actually feasible from \var{source}.

One last consideration is regarding the format of relations $\mathcal{C}_{?}$. Notice that in our chosen symbolic representation (BDDs), all elements have Boolean domains. Consequently, we cannot represent a control relation $\mathcal{C} \subseteq \bools_*^n \times \bools^m$ directly ($\bools_*^n$ is not a set of Boolean vectors). However, the result of \textsc{CanPerturb} already contains all information contained within $\mathcal{C}_{?}$ -- in the return statements, we thus simply describe a way of translating the information between $\mathcal{C}$ and the existing coloured vertex set \var{map}. As such, the \var{map} variable essentially represents one of the possible Boolean encodings of a non-Boolean relation $\mathcal{C} \subseteq \bools_*^n \times \bools^m$.

\section{Results}

To evaluate our methods, we have implemented algorithms from the previous section by using internal libraries of the tool AEON~\cite{aeon}. The prototype implementation is available as an open-source Github repository\footnote{\url{https://github.com/sybila/biodivine-pbn-control}}. Our experiments were run using a computer with AMD Ryzen Threadripper 2990WX 32-Core Processor and 64GB of memory. The results of experiments are discussed in Section~\ref{s:discussion}.

Our first experiment measures the performance of methods on five biological models with varying state space size (induced by the number of variables), a number of inputs, and a possible number of perturbations. All these three components influence the size of the coloured graph under consideration. Whereas the number of vertices is equal to the number of states, the number of all STG instances is equivalent to the product of colours and possible perturbations.

The first model is a \emph{myeloid} differentiation network \cite{Krumsiek2011} which models a bone marrow tissue cell differentiation from common myeloid cell to specialised blood cells (megakaryocytes, erythrocytes, granulocytes and monocytes). The second model depicts a heart development done by \emph{cardiac} kernel transcription factors~\cite{Waardenberg2014}. The third model explains regulatory interactions, which may cause \emph{ERBB} kinase over-expression -- a marker of breast cancer~\cite{Sahin2009}. The fourth model focuses on specific conditions which lead to a metastatic \emph{tumour}~\cite{Cohen2015}. The last model, the Mitogen-Activated Protein Kinase (\emph{MAPK}) network, consists of signalling pathways involved in diverse cellular processes, such as cell cycle, survival, apoptosis and differentiation~\cite{Grieco2013}. 
We derived partially specified versions of these models by substituting a portion of the network's update functions with uninterpreted functions of the same arity. This emulates the model development process, during which the general structure of the system is already determined, but the exact network dynamics are not yet fully established. All studied partially specified models can be found in the Github repository. 

We have selected two or three attractors from every fully specified model version obtained from the Cell Collective platform~\cite{Helikar2012}. We then use partially specified versions of these models. All algorithms are computed for all pairs of source and target attractors. For every model, using the introduced symbolic algorithms, we compute all types of control -- one-step, permanent, and temporary. For every computed attractor, we also state how big is the relevant colour space, as it is expected to have an influence on the computation time. The results are grouped by the target attractor, as the times for these results are similar given the nature of the algorithm (the most demanding part -- computation of target trap space depends solely on the choice of the target). The results can be found in Table~\ref{table:models_results}. This experiment demonstrates that our framework can handle even relatively large models (20 and more variables) in a convenient range of times. 

\begin{table}[hb!]
    \centering
    \caption{Performance of algorithms achieved on the biological models. Sub-tables correspond to particular models. The sub-table header displays the name of the model, reference to its source, and factors influencing the size of the underlying graph -- the number of states (induced by the number of the model variables), the approximate number of model colours, and employed types of control. The results are computed for some two or three arbitrary attractors. Therefore, for every target attractor, we list the number of relevant colours where the attractor is present and the average of times needed to compute one-step, permanent and temporary control from the source attractors to the said target attractor. Computation times marked as N/A did not finish in 4 hours.}
    
    \setlength\tabcolsep{2 pt}
      \renewcommand{\arraystretch}{1.1}
      \begin{tabular}{ | C{2.3cm} | C{2.3cm} | C{2.3cm} | C{2.3cm} | C{2.3cm} | }
            \hline
            \multicolumn{5}{|c|}{Myeloid~\cite{Krumsiek2011} -- $2^{11}$ states,  $\sim2^{17}$ colours, $2^{11}$ perturbations} \\ \hline
            Control target $t$ & Attract. colours & One-step & Permanent & Temporary \\
            \hline
            $A_1$ &  $\sim2^{10}$ & 8 ms & 15 ms & 10 ms \\
            $A_2$ &  $\sim2^{14}$ & 38 ms & 107 ms & 128 ms \\
            $A_3$ &  $\sim2^{16}$ & 62 ms & 124 ms & 607 ms \\
            \hline
            \multicolumn{5}{|c|}{Cardiac~\cite{Waardenberg2014} -- $2^{15}$ states, $\sim2^{28}$ colours, $2^{15}$ perturbations} \\ \hline
            Control target $t$ & Attract. colours & One-step & Permanent & Temporary \\
            \hline
            $A_1$ & $\sim2^{27}$ & 174 ms & 350 ms & 2 s \\
            $A_2$ & $\sim2^{27}$ & 520 ms & 2058 ms & 13 s \\
            $A_3$ & $\sim2^{27}$ & 453 ms & 1050 ms & 77 s \\
            \hline
            \multicolumn{5}{|c|}{ERBB~\cite{Sahin2009} -- $2^{20}$ states, $\sim2^{28}$ colours, $2^{20}$ perturbations} \\ \hline
            Control target $t$  &  Attract. colours & One-step & Permanent & Temporary \\
            \hline
            $A_1$ & $\sim2^{27}$ & 905 ms & 35 s & 8.2 min \\
            $A_2$ & $\sim2^{28}$ & 7141 ms & 40 s & 2 hrs \\
            \hline
            \multicolumn{5}{|c|}{Tumour~\cite{Cohen2015} -- $2^{32}$ states, $\sim2^{16}$ colours, $2^{32}$ perturbations} \\ \hline
            Control target $t$  &  Attract. colours & One-step & Permanent & Temporary \\
            \hline
            $A_1$ &  $\sim2^{14}$ & 1049 ms & 35 s & 8 min \\
            $A_2$ &  $\sim2^{14}$ & 887 ms  & 56 s & 15 min \\
            $A_3$ &  $\sim2^{14}$ & 947 ms  & 7.5 s & 20 min \\
            \hline
            \multicolumn{5}{|c|}{MAPK~\cite{Grieco2013} -- $2^{53}$ states,  $\sim2^{16}$ colours, $2^{53}$ perturbations} \\ \hline
            Control target $t$  &  Attract. colours & One-step & Permanent & Temporary\\
            \hline
            $A_1$ & $\sim2^{15}$ & 6.4 s & 3 hrs & N/A \\
            $A_2$ & $\sim2^{15}$ & 5.5 s & 1.2 hrs & N/A \\
            \hline
      \end{tabular}    
    \label{table:models_results}
\end{table}

The second demonstration inquires about the scalability of our algorithms with respect to the number of colours. The results are shown in Table~\ref{table:params_scal} for the tumour model~\cite{Cohen2015}. It can be seen that the times needed to compute the results grow much slower than the number of model colours. In particular, a jump from $2^{14}$ to $2^{21}$ colours, which represents a $128\times$ increase in the size of the colour space, results in an approx. $3-6\times$ increase in runtime.

\begin{table}[h!]
\centering
\caption{Algorithms scalability with respect to colours demonstrated on the tumour model~\cite{Cohen2015}. We selected one arbitrary target attractor in the model for all following computations. In the header, the number of colours containing an attractor with the computed target is listed. The shown time is the worst case of source-target control in all considered source attractors.}
            \renewcommand{\arraystretch}{1.3}

      \begin{tabular}{ | C{2.2cm} | C{1.8cm} | C{1.8cm} | C{1.8cm} | C{1.8cm} | C{2.0cm} |}

\hline

 \textbf{Rel. colours} & $\mathbf{2^0}$ & $\mathbf{\sim2^4}$ & $\mathbf{\sim2^{10}}$ & $\mathbf{\sim2^{14}}$ & $\mathbf{\sim2^{21}}$  \\ 
\hline
 \textbf{One-step} & 120 ms & 170 ms & 546 ms & 1053 ms & 3324 ms  \\ 
\hline
 \textbf{Permanent} & 3.2 s & 13.5 s & 14.2 s & 57.2 s & 262 s \\ 
\hline
 \textbf{Temporary} & 0.5 min & 1.3 min & 4.4 min & 20.4 min & 130 min  \\ 
\hline
\end{tabular}
\label{table:params_scal}

\end{table}

Next, we compare our one-step approach to the semi-symbolic one-step control method described in~\cite{Brim2021}. In the semi-symbolic method, only the BN inputs were specified symbolically, using BDDs. In contrast, using our fully symbolic approach presented here, the perturbations, as well as the sets of states, are represented symbolically, all by using BDDs. In Table~\ref{table:compare_semisymbolic}, we show a comparison of these two techniques. 

\begin{table}[!h]
    \centering
      \caption{Results of one-step control -- comparison of the fully symbolic approach to semi-symbolic approach presented in~\cite{Brim2021}. The values are stated
      as ranges because we computed one step-control on a set of attractors. The first column contains name of model and number of its states. The second column shows the number of model's colours. The third column shows count ranges of
      colours which contain the given target attractor. The fourth and fifth columns display ranges of times to compute one-step controls (strong basins) using semi-symbolic respectively symbolic methods. The computation time marked as N/A was not feasible to compute due to insufficient memory.}
      \centering
      \setlength{\tabcolsep}{0.5em} 
      \renewcommand{\arraystretch}{1.3}
      \begin{tabular}{|c|c|c|c|c|}
            \hline
            \textbf{Model} & \textbf{Colours} & \textbf{Attract. colours} &  \textbf{Semi-symbolic} & \textbf{Symbolic} \\
            \hline
            \multirow{3}{*}{\begin{tabular}{c} Cell-Fate~\cite{Calzone2010} \\ $2^{19}$ states\end{tabular}} & 1 & 1 & 4.4 -- 9.19 s & 5 -- 21 ms  \\ 
            & $\sim2^{10}$ & 1 -- 4 & 4.63 -- 13.42 s & 103 -- 317 ms \\ 
            & $\sim2^{24}$ & 7 -- 56 & 5.82 -- 26.59 s & 3 -- 17.4 s \\ 
            \hline
            \multirow{3}{*}{\begin{tabular}{c} Myeloid~\cite{Krumsiek2011} \\ $2^{11}$ states\end{tabular}} & 1 & 1 &  8 -- 30 ms & $\leq$ 1 ms\\ 
             & $\sim2^{26}$ & 63 -- 2,052 & 14 -- 214 ms & 43 -- 460ms\\ 
          & $\sim2^{49} $ & $\sim2^{16}$ -- $\sim2^{24}$  & 147 -- 1717 ms & 1.4 -- 14 min \\ 
            \hline
            \begin{tabular}{c} MAPK~\cite{Grieco2013} \\ $2^{53}$ states\end{tabular} & $\sim2^{16}$ & $\sim2^{15}$ & N/A & 5.5 -- 6.4 s 

            \\
            \hline
      \end{tabular}
      \label{table:compare_semisymbolic}
\end{table}

\begin{figure}[h!]
\caption{Maximal robustness $\rho_{max}$ per set of perturbations with a given size -- comparison between different types of controls. The $x$-axis shows the size of perturbations in the set and the $y$-axis displays the maximal robustness. Each chart is showing robustness for a different model.}
\centering
\begin{tikzpicture}[scale=0.65, transform shape]
\begin{axis}[\axisStyle{a) Cardiac (15 variables)}{Maximal}]
    \addplot coordinates {(1,0.095)(2,0.393)(3,0.404)(4,0.599)(5,0.696)};
    \addplot coordinates {(1,0.095)(2,0.685)(3,0.841)(4,0.987)(5,1.0)};
    \addplot coordinates {(1,0.095)(2,0.685)(3,0.841)(4,0.987)(5,1.0)};
    \legend{One-step,Permanent,Temporary}
\end{axis}
\end{tikzpicture}
\begin{tikzpicture}[scale=0.65, transform shape]
\begin{axis}[\axisStyle{b) ERBB (20 variables)}{Maximal}]
\addplot coordinates {(1,0.034)(2,0.035)(3,0.036)(4,0.215)(5,0.240)};
\addplot coordinates{(1,0.215)(2,0.621)(3,0.862)(4,1.0)(5,1.0)};
\addplot coordinates{(1,0.215)(2,0.621)(3,0.862)(4,1.0)(5,1.0)};
\legend{One-step,Permanent,Temporary}
\end{axis}
\end{tikzpicture}
\begin{tikzpicture}[scale=0.65, transform shape]
\begin{axis}[\axisStyle{c) Tumour (32 variables)}{Maximal}]
\addplot coordinates {(1,0.169)(2,0.197)(3,0.209)(4,0.483)(5,0.563)};
\addplot coordinates{(1,0.434)(2,0.652)(3,0.869)(4,1.0)(5,1.0)};
\addplot coordinates{(1,0.434)(2,0.652)(3,0.869)(4,1.0)(5,1.0)};
\legend{One-step,Permanent,Temporary}
\end{axis}
\end{tikzpicture}
\label{fig:maximal_robustness}
\end{figure}

\begin{figure}[h!]
\caption{Union robustness $\rho_{max}$ per set of perturbations with a given size -- comparison between different types of controls. The $x$-axis shows the size of perturbations in the set and the $y$-axis displays the maximal robustness. Each chart is showing robustness for a different model.}
\centering
\begin{tikzpicture}[scale=0.65, transform shape]
\begin{axis}[\axisStyle{a) Cardiac (15 variables)}{Union}]
\addplot coordinates {(1,0.095)(2,0.395)(3,0.414)(4,0.649)(5,0.743)};
\addplot coordinates {(1,0.095)(2,0.674)(3,0.962)(4,0.999)(5,1.0)};
\addplot coordinates {(1,0.095)(2,0.792)(3,0.986)(4,0.999)(5,1.0)};
\legend{One-step,Permanent,Temporary}
\end{axis}
\end{tikzpicture}
\begin{tikzpicture}[scale=0.65, transform shape]
\begin{axis}[\axisStyle{b) ERBB (20 variables)}{Union}]
\addplot coordinates {(1,0.035)(2,0.036)(3,0.036)(4,0.220)(5,0.252)};
\addplot coordinates {(1,0.220)(2,0.806)(3,0.985)(4,1.0)(5,1.0)};
\addplot coordinates {(1,0.220)(2,0.806)(3,0.985)(4,1.0)(5,1.0)};
\legend{One-step,Permanent,Temporary}
 \end{axis}
\end{tikzpicture}
\begin{tikzpicture}[scale=0.65, transform shape]
\begin{axis}[\axisStyle{c) Tumour (32 variables)}{Union}]
\addplot coordinates {(1,0.198)(2,0.208)(3,0.209)(4,0.523)(5,0.610)};
\addplot coordinates {(1,0.576)(2,0.902)(3,0.991)(4,1.0)(5,1.0)};
\addplot coordinates {(1,0.576)(2,0.902)(3,0.991)(4,1.0)(5,1.0)};
\legend{One-step,Permanent,Temporary}
 \end{axis}
\end{tikzpicture}
\label{fig:total_robustness}
\end{figure}

Finally, we conduct experiments evaluating the presented robustness metrics across different types of controls. In particular, we use maximal robustness $\rho_{max}$ and union robustness $\rho_{union}$ metrics which we measure on sets of fixed perturbation sizes. The results are shown in Fig.~\ref{fig:maximal_robustness} and Fig.~\ref{fig:total_robustness}. The models are the same for both experiments and the same as in our first experiment (see Table~\ref{table:models_results}). For demonstration purposes, an arbitrary but fixed source and target attractor have been selected among the ones considered in Table~\ref{table:models_results}. 

\section{Discussion}
\label{s:discussion}

In the presented work, we show how to employ symbolic methods to solve the control of partially specified BNs. The symbolic representation is used to encode three types of entities: a set of the STG vertices, a set of colours (input valuations) which admit an edge in the STG, and a set of perturbations that permit the given edge. When we increase the number of colours (unknown parts of the system), only one dimension of the resulting symbolic representation grows. In opposition, when we increase the number of system's variables, both the size of the STG, and the available perturbations grow. Therefore, it is expected that increasing the model size has a bigger impact on the performance than increasing the number of model's parameters. This hypothesis is to some extent supported by our two experiments captured in Table~\ref{table:models_results} and Table~\ref{table:params_scal}. We can observe that when the colour space increases in isolation, the model still remains processable and the computation time grows much slower than the absolute number of colours. On the other hand, increasing the size of the state space can result in analysis being infeasible (see the MAPK case).

Nonetheless, it must be highlighted that in both of these cases, the impact of scaling up colours versus variables is to some extent unpredictable. This is given by the intrinsic mechanics of the introduced STGs. As the nature of our method relies on fixed-point computation, such that each iteration eliminates only the direct predecessors of a particular set, in cases where e.g. a long path in the graph arises, our method will not be able to scale well: the algorithm will need many iterations to converge, each eliminating only a small fraction of the graph vertices. Another unforeseeable effect on the method's performance have the BDDs themselves. The ordering of the BDD diagram has a significant impact on the effectiveness of all used operations and no single ordering strategy can perform best in all cases~\cite{Harlow2001}.

Another insight that can be obtained from Table~\ref{table:models_results} is how one-step, temporary and permanent controls differ in their complexity in terms of computational time. We can notice that one-step control is significantly over-performing other methods. This is because, in the case of one-step control, we do not need to consider perturbation state-space when computing the trap space of the target attractor. In contrast, the temporary control is by far the most computationally demanding compared to the other types control. This is again given by the fact that we need to compute two distinct trap spaces. On the one hand, the first computed trap space is the same as in the case of one-step control and should be thus easy to compute (the perturbation space is irrelevant). However, on the other hand, the seed for the second computed trap space is a much bigger vertex set, and therefore the resulting second trap space (which includes perturbations) of the first trap space (without perturbations) can generate a much more complex symbolic structure.

Apart from the time-wise performance, we also compared the robustness of different types of control (see Figure~\ref{fig:maximal_robustness} and Figure~\ref{fig:total_robustness}). At first, let us look at what particular types of robustness can tell us about the inspected systems. Under the assumption that the real biological system corresponds to one of the instantiations of our partially specified BN model, what the maximal robustness tells us is how high is the chance that \emph{a particular perturbation} will work in the real model. Alternatively, we use union robustness to evaluate a chance that there is \emph{some perturbation} of the given size that will work. For example, in the ERBB model (Figure~\ref{fig:maximal_robustness}b and Figure~\ref{fig:total_robustness}b), we can notice that there is a 60\% chance that the most robust perturbation (temporary or permanent) of size two will work for the real, fully specified model. However, even if this particular perturbation fails, we know that 80\% of instantiations can be influenced by a perturbation of this size. This means that out of the 40\% of instantiations that cannot be influenced by the perturbation with the maximal robustness, a half is still controllable by perturbing only two variables. For all three models, it is thus highly likely that a temporary perturbation of size no bigger than 3 exists which controls the network. That is a very small portion of the total model variables which could be possibly perturbed.

As the one-step control is a special case of the temporary control, it is expected that temporary perturbations are achieving the same or higher robustness than the one-step case. This is also demonstrated in the experiments, where we find no instance in which the one-step control could achieve a higher maximal or union robustness. We can also observe that temporary and permanent controls mostly perform similarly well. This is in line with~\cite{Su2019FM} where the minimal temporary and permanent controls also performed the same in the most of the cases. In our case, only in the cardiac model, the temporary control outperforms the permanent one in the union robustness (Figure~\ref{fig:total_robustness}a). This means there are some perturbations which work when applied temporarily as opposed when applied permanently. There are two admissible explanations for this. The first may be a simple fact that the trap space computed for temporary control is expected to be bigger, since the seed of the trap space is not only the given target attractor, but its trap space in an unperturbed case. The second explanations for this is a scenario when a permanent perturbation disrupts the BN dynamics in such a way that the target attractor completely ceases to exist. 

Another insight of our experiments is a comparison of the novel symbolic approach to the semi-symbolic approach from~\cite{Brim2021}. The experiment captured in Table~\ref{table:compare_semisymbolic} shows that the symbolic approach is performing better on larger models with fewer colours. Nonetheless, if the model is small but has many colours, the semi-symbolic approach can outperform the fully symbolic one. This is understandable due to the already discussed heuristic nature of BDDs: in some cases, the ordering of BDD nodes might not be optimal, resulting in large underlying memory consumption and longer computation. Nonetheless, the greatest benefit of the novel symbolic approach lies in the ability to process models with a much bigger state space. This is the case of the MAPK model, where the explicit state-space of the semi-symbolic approach completely fails to fit into the memory.

Finally, we compare our results with the related work discussed in Section~\ref{s:introduction}. It is important to note that the previously published methods other than~\cite{Brim2021} do not support partially specified BNs. Therefore, we deem our method incomparable to the previous line of research in terms of performance. Hypothetically, the other methods could be used to iteratively process all instantiations of a partially specified BN, resulting in a na\"ive parameter-scan approach. Nonetheless, as the number of instantiations grows exponentially, this approach cannot scale in the same way and thus quickly become infeasible. More than that, our approach allows easy evaluation of robustness, which would otherwise also be more complicated when the results are computed na\"ively. Considering other aspects of BN control, our method supports all types of static perturbations (one-step, temporary, permanent)~\cite{Su2020}. However, it does not support more complex, dynamical application of sequential perturbations~\cite{Mandon2019CMSB,Su2020}. With that type of perturbations it is possible to achieve the minimal perturbation of a very small size. However, we consider such approach too computationally demanding for the case of partially specified BNs at the moment -- the combinatorial space of available perturbations, colours, and attractors (each present only for a subset of colours) would be enormous. Finally, our method supports only source-target control. We leave other types of control, e.g., the target control~\cite{JunPang2020a} and the phenotype control~\cite{Fontanals2021} for future work.

\section{Conclusion}

In this work, we have developed methods and efficient algorithms to solve the source-target control problem for partially specified BNs using one-step, temporary and permanent state perturbations. The methods we proposed are entirely symbolic and are based on BDDs. We demonstrated that our approach is capable of controlling models with a high degree of uncertainty in a conveniently short time. Such a result
cannot be achieved with the na\"ive scan over all BN inputs due to the doubly exponential explosion of the number of the inputs representing the fully specified BN instances. We also showed that our novel methods are superior to the existing semi-symbolic one-step control approach~\cite{Brim2021} in terms of model size we can process as well as the robustness of temporary and permanent perturbations. We were able to evaluate robustness between different types of control thanks to the metrics we introduced -- maximal and union robustness.

In future work, we would like to improve the algorithms for computing
the strong basin, e.g., by employing suitable network
decomposition procedures~\cite{Mizera2018,Su2019TCBB}. We would also like to allow target control or only a partial specification of the state space into which the system is controlled (i.e. trap spaces or phenotypes). Another possible extension to all algorithms
would be to implement a restriction on the control size or on variables that may be perturbed. Next, we aim to explore
practical applications of the developed methods further. If these inquiries show that permanent perturbations are practically applicable, we might study further this type of perturbations and its potential application for phenotypes variance. Last but not least,
we plan to incorporate the control methods into our
toolkit for partially specified BNs -- the AEON tool~\cite{aeon}.

\bibliographystyle{unsrt}  
\bibliography{references}  


\end{document}